\documentclass[journal]{IEEEtran}

\usepackage{cite}
\usepackage{amsmath}
\usepackage{amsthm}
\usepackage{algorithmic}
\usepackage{url}

\usepackage{amsfonts}
\usepackage{color}
\usepackage{graphicx}

\usepackage{pstricks}
\usepackage{pst-plot}
\usepackage{pst-node}
\usepackage{psfrag}
\usepackage{url}
\usepackage{algorithmic}
\usepackage{algorithm}

\graphicspath{{./}}
\DeclareGraphicsExtensions{.eps}

\newcommand{\fig}{Figure\,}
\newcommand{\tab}{Table\,}
\newcommand{\sect}{Section\,}

\newcommand{\cf}{cf.\,}
\newcommand{\ie}{i.e.\,}

\newcommand{\degc}{^{\circ}\text{C}}

\newcommand{\R}[1]{\mathbb{R}^{#1}}
\newcommand{\set}[1]{\mathbb{#1}}
\newcommand{\e}{\begin{equation}}
\newcommand{\ee}{\end{equation}}
\newcommand{\opt}{^{\ast}} 
\newcommand{\ind}[1]{^{(#1)}}

\newcommand{\ts}[1]{\text{#1}}

\newcommand{\pr}{p_r}

\newcommand{\Tout}{\theta_{\ts{out}}}
\newcommand{\Tin}{\theta_{\ts{in}}}

\begin{document}

%
\title{Large-Scale Demonstration of Precise Demand Response Provided by Residential Heating Systems}
%
%
%

\author{Fabian~L.~M\"uller \&
		Bernhard~Jansen
%
\thanks{This work was supported by the Danish Energy Agency's Energy Technology Development and Demonstration Program under the project \textsc{Ecogrid2.0}.}
\thanks{F. L. M\"uller is with the Automatic Control Laboratory, Swiss Federal Institute of Technology, Zurich, Switzerland, and IBM Research--Zurich, Rueschlikon, Switzerland.
        {\tt\small fmu@zurich.ibm.com}}%
\thanks{B. Jansen is with IBM Research--Zurich, Rueschlikon, Switzerland.
        {\tt\small bja@zurich.ibm.com}}%
}

\maketitle

\begin{abstract}
Being able to adjust the demand of electricity can be an effective means for power system operators to compensate fluctuating renewable generation, to avoid grid congestion, and to cope with other contingencies. Electric heating and cooling systems of buildings can provide different demand response services because their electricity consumption is inherently flexible because of their thermal inertia. 
This paper reports on the results of a large-scale demand response demonstration involving a population of more than 300 residential buildings with heat pump installations. We show how the energetic behavior and flexibility of individual systems can be identified autonomously based only on energy meter data and outdoor air temperature measurements, and how the aggregate demand response potential of the  population can be quantified. Various load reduction and rebound damping experiments illustrate the effectiveness of the approach: the load reductions can be predicted precisely and amount to 40--65\% of the aggregate load, and the rebound can be damped efficiently.

\end{abstract}

\begin{IEEEkeywords}
Demand response, direct load control, energetic flexibility, thermostatically controlled loads
\end{IEEEkeywords}

\section{Introduction}
\label{s:introduction}
\IEEEPARstart{K}{eeping} the electricity demand and supply in a power system balanced at all times can be challenging for system operators (SO), in particular in the light of growing shares of uncertain and intermittent renewable generation \cite{GFR2017}. Demand response (DR) is one approach proposed to help balance the power grid: it aims to schedule and adjust the electricity consumption of systems intelligently according to current supply or to meet certain grid requirements. DR schemes have been studied extensively in theory, and there exist various projects implementing and evaluating different DR mechanisms. Comprehensive overviews of past and current DR projects can be found in \cite{Gangale2017,GridInnovation,Obinna2016}. 

DR schemes can be sorted into two main categories, namely, direct control and indirect control. 
%
Indirect control refers to a setup in which an incentive signal, such as a forecast of the electricity price, is broadcast to the consumers, which are expected to adjust their demand accordingly. The fact that only a single incentive signal needs to be broadcast and that control of the consumer devices is the responsibility of their operators makes indirect load control a simple, privacy-preserving and highly scalable approach for SOs. However, the demand sensitivity with regard to the incentive signal is unknown and must be estimated.
Implementations of indirect load control schemes considering buildings are, among others, the \textsc{EcoGridEU} project, in which 1800 residential heat pumps (HP) and electric heaters reacted to a price signal \cite{Liu2016}. 
Different DR experiments were conducted in the \textsc{Linear} \cite{Linear2014}, the \textsc{Adress} \cite{Eyrolles2013}, and the \textsc{Advanced} \cite{Advanced2014} projects involving 460, 400, and 300 household appliances, respectively. 
Other examples of projects are \textsc{gridSmart} \cite{Widergren2014}, \textsc{Grid4EU} \cite{Grid4eu2016}, \textsc{Olympic Peninsula} \cite{Chassin2008}, and \cite{Kawamura2016,Klaassen2016,Bliek2010}.

Direct load control, in contrast, refers to the case in which loads are controlled directly via a control signal that is applied to individual systems. In this setup, an aggregator (AG) controls the electricity consumption of one or several systems, identifies their energetic flexibility, and offers DR services to the SO. The AG is responsible to meet each system's energy needs while satisfying both operational and comfort constraints. Direct control requires two-way communication between the AG and each resource to send control commands and receive feedback on the system state. The main advantage of direct control is that it allows the AG to coordinate the DR of a group of systems, making it an accurate and versatile method to provide different types of services \cite{Callaway2011}. 
Examples of direct control DR projects are, among others, the \textsc{moma} project, which involved 73 buildings and achieved load shifts of 6--8\% of the aggregate demand \cite{Moma2013}. The ability to track power references by a population of 54 HPs was illustrated in \cite{Biegel2014}. The \textsc{Hartley Bay} project achieved a maximum load reduction of 36 kW from controlling 32 residential heating and cooling devices \cite{Wrinch2012}. The results in \cite{DeSomer2017} show that controlling the hot water buffers of 6 buildings could increase their photovoltaic self-consumption by more than 20\%.

This work deals with direct load control and considers the tasks of an AG controlling a population of residential HPs with the goal of providing load reduction services to the SO. The AG must identify the energetic behavior of all individual systems and characterize their flexibility with regard to providing the DR service. Different approaches have been used to characterize the DR behavior of groups of systems. Top-down approaches attempt to capture the aggregate behavior directly based, for example, on the probabilistic properties of the underlying systems \cite{Kamgarpour2013,Sajjad2016,Callaway2009,Zhang2012a}. In contrast, we consider a bottom-up approach in which the behavior of each system is described first and aggregated subsequently, see \cite{Hao2016,Mathieu2015,DeConinck2013,Sanandaji2013}, among others. Bottom-up approaches require knowledge about the physical parameters and constraints of individual systems, which, in general, are unknown to the AG.
In particular for large populations of resources, it usually is prohibitively expensive and time-consuming to collect the nameplate and measurement data required by classical system identification techniques. 



%

Our main contributions are twofold. First, we present a characterization of the aggregate DR behavior of a population of buildings explicitly taking into account the thermal properties and constraints of individual systems. Our approach requires only limited, readily available measurement data and can easily be automated, making it an inexpensive and highly scalable tool for AGs. Second, we describe a large-scale, real-life implementation of a direct-control DR scheme, discuss our key findings, and present the results from various DR experiments involving more than 300 residential buildings to prove the effectiveness of our approach.


The paper is organized as follows. \sect\ref{s:expSetup} describes the experimental setup. \sect\ref{s:sysModeling} introduces a model of the energy dynamics of a building and its heating system, whose key parameters are identified  in \sect\ref{s:sysIdent}. In \sect\ref{s:popModeling}, the aggregate DR behavior of a population of systems is characterized. Experimental results are presented in \sect\ref{s:results}, and conclusions are provided in \sect\ref{s:conclusion}.
\section{Experimental Setup}
\label{s:expSetup}
\begin{figure}[tbh]
\centering 
\resizebox{0.48\textwidth}{!}{
\includegraphics[]{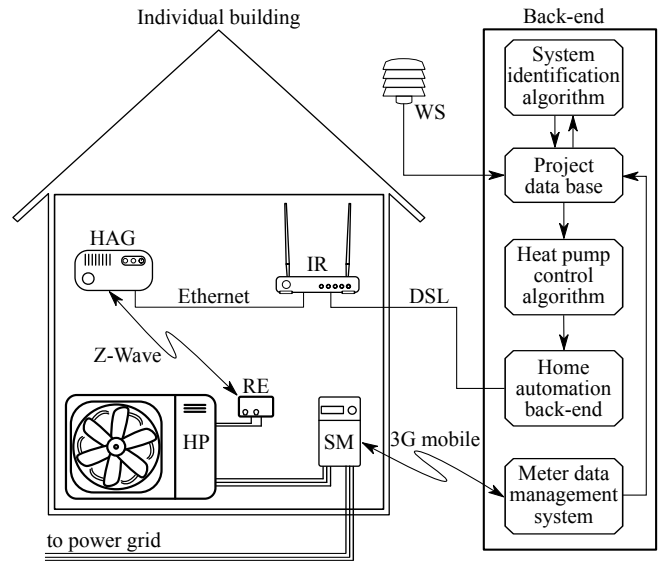}
}
\caption{Experimental setup used in this work: Each building is equipped with a smart meter (SM) that measures the energy consumption and generation of the entire building on a timescale of 5 min. The data is sent to the meter data management system and pushed to the project data base, which also hosts outdoor air temperature data provided by a weather service (WS). The system identification uses this data to estimate the energetic behavior and flexibility of a building. A control algorithm can send throttle commands to the HP via the home automation back-end and the home automation gateway (HAG).}
\label{fig:expSetup} 
\end{figure}

The experimental setup used in this work was originally established during the \textsc{EcogridEU} project \cite{Liu2016} and is reused in the successor project \textsc{Ecogrid2.0} \cite{ecogrid2} considered here. The setup comprises more than 300 inhabited residential buildings on the Danish island of Bornholm. \fig\ref{fig:expSetup} provides an overview of the experimental setup. The buildings are of different size, feature distinct thermal properties, and are equipped with HPs of different makes, types, and age.
%
%
In the \textsc{EcogridEU} project, each building was equipped with an off-the-shelf \emph{Landis+Gyr} \emph{E450} or \emph{E650} smart energy meter (SM) that samples the total active and reactive electric energy consumption and production (photovoltaic or wind) of the building with a sampling time of 5 min rather than the 15 min or 60 min intervals commonly used. Once a day, the meters upload their measurement data via a mobile internet connection to a central meter data management system by \emph{Landis+Gyr}, from where the data is pushed to the project data base. The latter also stores outdoor air temperature values measured at a single location on the island as well as corresponding forecasts provided by a weather service (WS). The system identification procedure introduced in \sect\ref{s:sysIdent} estimates the energetic behavior and flexibility of a building and is based exclusively on these measurement data. The 5-min energy data proved crucial for this purpose. The system identification is executed only once for each building, and the results are stored in the data base.
The HP control algorithm relies on the system properties identified when controlling the HP via a binary \emph{throttling signal}: If a value of 1 is sent to a HP that is currently running, a rundown sequence is initiated, and the HP will cease operation within 2--10 min depending on its current state. As long as the signal remains 1, the HP cannot turn on. However, once the throttling signal is released, \ie, is reset to 0, the HP can operate freely according to its internal thermostat controller (TC). Thus, the throttling signal under our control can be interpreted as a ``request to turn off'' if set to 1, and as a ``permit to operate according to TC'' otherwise. It is important to note that with this control setup it is impossible to force a HP to turn on and consume power.  

The control algorithm cannot access the HP directly but sends the throttling signal to the home automation back-end, which communicates with the HP via several stations: It uses a DSL connection to the internet router of the building to access the home automation gateway (HAG) via Ethernet. The HAG controls a \emph{Danfoss RXZ-1} relay (RE) via \emph{Z-Wave} to apply the throttling signal to the HP. To do so, it uses the so-called \emph{tariff input}, which is a feature common to all our HPs that can be used by the Distribution System Operator to prevent the HPs from consuming electricity in situations where there is a risk of distribution grid overload. 

The multi-state communication between the control algorithm and the HP proved to be inconvenient and unreliable for several reasons. First, finding a suitable position for the HAG can be tricky because it requires a cable connection to the IR but must at the same time be located such that the RE is reachable via \emph{Z-Wave} radio signal. Power-line Ethernet bridges have been used to relax these restrictions. Second, the setup involves at least three on-premise devices, namely, the IR, HAG, and RE. If one of them is unplugged or powered off the control algorithm can no longer communicate with the HP. To prevent prolonged throttling of a HP in case of communication problems, the control algorithm always sends a throttling request together with a predefined throttling release time, which is stored locally on the HAG and is applied to the RE even if the communication between the control algorithm and the HAG breaks down.
%
%
%
%
To make a future communication and control setup leaner and more reliable for DR purposes, we suggest to use the digital output of the SM, which is provided by many SM models including the \emph{Landis+Gyr} meters used here, to control the RE directly.
%
%
%
\vspace*{5mm}
\vfill
\section{System Modeling}
\label{s:sysModeling}


An individual building and its heating system are modeled as a single energy buffer that represents the storage of thermal energy in the building as well as in the heating circuit of the HP. The evolution of the building's energy level ${e(t)\in\R{+}}$ is governed by the in- and out-flux of thermal energy:
\begin{equation}
	de(t)/dt=q_{c}(t)u(t)-q_{l}(t),
	\label{eq:energyDyn}
\end{equation}
where ${q_{c}(t),\,q_{l}(t)\in\R{}}$ are referred to as the  energy \emph{charge} and \emph{loss} rates, respectively. For the sake of simplicity, we only consider the two most common operational states of the HP, ON and OFF, as represented by the binary control input ${u(t)\in\{1,0\}}$, respectively. Other modes such as electrical boosting, defrosting, or hot water production, are neglected. 
%
The charge rate equals the rate at which the HP feeds thermal energy into the system:
\begin{equation}
	q_c(t) = \eta(\cdot)\pr,
	\label{eq:chargeRate}
\end{equation} 
where $\pr$ is the rated power of the HP, and $\eta(\cdot)$ is its coefficient of performance (COP) that can depend on various factors, \cf\sect\ref{ss:identChargeRate}. 
%
The energy loss rate is assumed to depend linearly on the in- and outdoor air temperature $\theta_{\ts{in}}(t)$ and $\theta_{\ts{out}}(t)$, respectively:
\begin{equation}
	q_l(t) = (\theta_{\ts{in}}(t)-\theta_{\ts{out}}(t))/R,
	\label{eq:lossRate}
\end{equation}
where ${R\in\R{}}$ is a lumped parameter incorporating the different types of heat-transfer coefficients involved. 

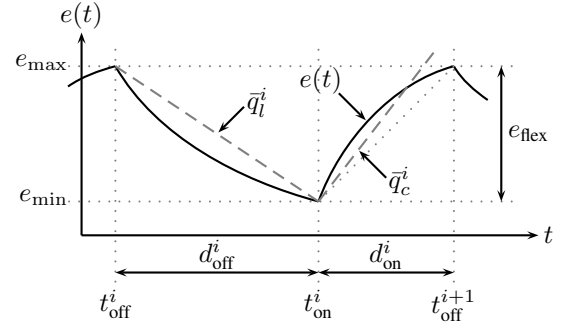
\begin{figure}[tb]
\setlength{\abovecaptionskip}{-0pt}
\centering
\psset{unit=0.9}
\begin{pspicture}(-0.75,-1.25)(7,3.6)
	
	\psline[]{->}(0,0)(6.75,0)
	\psline[]{->}(0,0)(0,3)
	\rput[c](6.9,0){$t$}
	\rput[b](0,3.05){$e(t)$}
	
	\psline[linestyle=dotted,linecolor=gray]{}(-0.2,2.5)(6.5,2.5)
	\psline[linestyle=dotted,linecolor=gray]{}(-0.2,0.5)(6.5,0.5)
	\rput[Br](-0.2,0.5){$e_{\min}$}
	\rput[Br](-0.2,2.5){$e_{\max}$}
	
	\psline[linestyle=dotted,linecolor=gray]{}(0.5,-0.7)(0.5,2.75)
	\psline[linestyle=dotted,linecolor=gray]{}(3.5,-0.7)(3.5,2.75)
	\psline[linestyle=dotted,linecolor=gray]{}(5.5,-0.7)(5.5,2.75)
	\rput[t](0.5,-0.8){$t_{\ts{off}}^i$}
	\rput[t](3.5,-0.8){$t_{\ts{on}}^i$}
	\rput[t](5.5,-0.8){$t_{\ts{off}}^{i+1}$}
	
	\pnode(-0.2,2.2){A}
	\pnode(0.5,2.5){B}
	\pnode(3.5,0.5){C}
	\pnode(5.5,2.5){D}
	\pnode(6,2){E}
	\nccurve[angleA=35,angleB=-165]{A}{B}
	\nccurve[angleA=-60,angleB=165]{B}{C}
	\nccurve[angleA=70,angleB=-165]{C}{D}
	\nccurve[angleA=-60,angleB=145]{D}{E}
	
	\psline[linestyle=dashed,linecolor=gray]{}(0.5,2.5)(3.5,0.5)
	\psline[linestyle=dotted,linecolor=gray]{}(3.5,0.5)(5.5,2.5)
	\psline[linestyle=dashed,linecolor=gray]{}(3.5,0.5)(5.2,2.7)

	\psline[]{<->}(0.5,-0.55)(3.5,-0.55)
	\psline[]{<->}(3.5,-0.55)(5.5,-0.55)
	\rput[b](2,-0.5){$d_{\ts{off}}^i$}
	\rput[b](4.5,-0.5){$d_{\ts{on}}^i$}
	
	\rput[b](2.6,1.75){$\bar{q}_l^i$}
	\psline[]{<-}(2.02,1.52)(2.4,1.9)
	\rput[b](4.7,0.6){$\bar{q}_c^i$}
	\psline[]{<-}(4.1,1.25)(4.5,0.85)
	
	\psline[]{<->}(6.2,0.5)(6.2,2.5)
	\rput[l](6.3,1.5){$e_{\ts{flex}}$}
	\psline[]{->}(3.8,2.1)(4.2,1.7)
	\rput[br](3.85,2.1){$e(t)$}
	
\end{pspicture}
\caption{Evolution of the thermal energy content $e(t)$ (solid) of a system with thermostatically controlled heating. The TC observes the upper and lower energy limits, $e_{\max}$ and $e_{\min}$. On- and off-switching times of the HP are denoted by $t_{\ts{on}}^i$ and $t_{\ts{off}}^i$, with corresponding on- and off-durations $d_{\ts{on}}^i$ and $d_{\ts{off}}^i$ for heating cycle $i$. Also shown are the average charge and loss rates $\bar{q}_c^i$ and $\bar{q}_l^i$ (dashed).} 
\label{fig:energyDynamics}
\end{figure}

The HP is controlled by a TC which switches the HP on (off) as soon as the energy level $e(t)$ reaches the user-defined lower (upper) bound $e_{\min}$ ($e_{\max}$). Thus, the amount of flexible energy available during normal operation is ${e_{\ts{flex}}:=e_{\max}-e_{\min}}$. \fig\ref{fig:energyDynamics} shows the energy trajectory resulting from this type of control.
The on- and off-durations of the heating cycle $i$ are denoted by $d_{\ts{on}}^i$ and $d_{\ts{off}}^i$, respectively.
The energy dynamics \eqref{eq:energyDyn} depend on the internal state of the system as reflected by the indoor air temperature $\theta_{\ts{in}}(t)$ in \eqref{eq:lossRate}. However, because the TC keeps $\theta_{\ts{in}}(t)$ within an interval that is narrow compared with ${\theta_{\ts{in}}(t)-\theta_{\ts{out}}(t)}$, we assume that the charge and loss rates $q_c(t)$ and $q_l(t)$ can be approximated for every heating cycle $i$ by their average values $\bar{q}_c^i$ and $\bar{q}_l^i$, \ie,
\begin{align}
	q_c(t)=\bar{q}_c^i &:= \left(\frac{1}{d_{\ts{on}}^i}+\frac{1}{d_{\ts{off}}^i}\right) e_{\ts{flex}}, &
	q_l(t)=\bar{q}_l^i &:= \frac{e_{\ts{flex}}}{d_{\ts{off}}^i},
	\label{eq:defAvgRates}
\end{align}
for ${t\in [t_{\ts{off}}^i,\,t_{\ts{off}}^{i+1})}$. The resulting energy dynamics are
\begin{equation}
	de(t)/dt = \bar{q}_c^i u(t)-\bar{q}_l^i,\ t\in[t_{\ts{off}}^i,\,t_{\ts{off}}^{i+1}).
	\label{eq:approxEnergyDyn}
\end{equation}
Because we are unable to measure the absolute amount of thermal energy $e(t)$ stored in the system, we consider the normalized state ${x(t):=(e(t)-e_{\min})/(e_{\max}-e_{\min})}$. From ${e_{\min}\leq e(t)\leq e_{\max}}$ guaranteed by the TC follows that ${0\leq x(t)\leq 1}$. Thus, $x(t)$ can be interpreted as the ``state of charge'' of the system and evolves according to
\begin{equation}
	dx(t)/dt = \bar{r}_c^i u(t) - \bar{r}_l^i,\,\ t\in[t_{\ts{off}}^i,\,t_{\ts{off}}^{i+1}),
	\label{eq:socDyn}
\end{equation}
with the \emph{normalized} average charge and loss rates
\begin{align}
	\bar{r}_c^i &:= \frac{1}{d_{\ts{on}}^i}+\frac{1}{d_{\ts{off}}^i} & \ts{and} & & 
	\bar{r}_l^i &:= \frac{1}{d_{\ts{off}}^i}.
	\label{eq:defNormRates}
\end{align}

The main advantage of the dynamics \eqref{eq:socDyn}--\eqref{eq:defNormRates} is that for each heating cycle $i$ they are fully defined by the corresponding on- and off-durations $d_{\ts{on}}^i$ and $d_{\ts{off}}^i$, respectively, together with the initial state ${x(t_{\ts{off}}^i)=1}$. 

\section{System Identification}
\label{s:sysIdent}

\subsection{Estimation of heat pump state and rated power}
\label{ss:identRatedPower}

\begin{figure}[tb]
\setlength{\abovecaptionskip}{-5pt}
\centering
\psfrag{xl2}[Bc][Bc][1.3]{Time}
\psfrag{yl1}[Bc][Bc][1.3]{$p_k$ kW}
\psfrag{yl2}[Bc][Bc][1.3]{$\Delta_2p_k$ kW}
\psfrag{a}[Bl][Bl][1.3]{$\Delta_{2,\ts{off}}$}
\psfrag{b}[Bl][Bl][1.3]{$\Delta_{2,\ts{on}}$}
\resizebox{0.495\textwidth}{!}{
\includegraphics{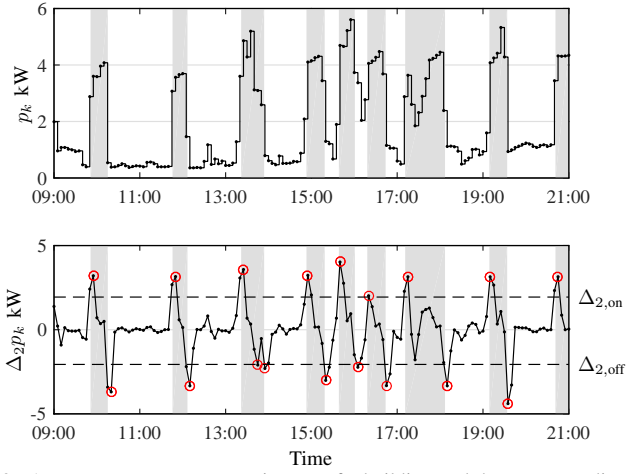}
}
\caption{Aggregate power consumption $p_k$ of a building and the corresponding changes over two subsequent time steps $\Delta_2 p_k$. The thresholds $\Delta_{2,\ts{on}}$ and $\Delta_{2,\ts{off}}$ are used to detect potential HP switching events. The shaded gray areas denote time periods during which the HP is believed to be ON.}
\label{fig:powerSteps}
\end{figure}

The SM measures the total cumulative energy consumption $e_k$ of a building with a sampling time of ${t_s=5}$ min, where ${k}$ indexes discrete time. These energy measurements are translated into power values
\begin{equation}
	p_k:=(e_{k+1}-e_k)/t_s,\ k=0,\dots,N,
	\label{eq:defPower}
\end{equation}
where $p_k$ is interpreted as the \emph{average} power consumption during the time interval ${[kt_s,(k+1)t_s)}$. A typical load profile of a building with predominant HP consumption is shown in the top plot in \fig\ref{fig:powerSteps}. 
Changes of the operational state of the HP usually involve significant changes in its power consumption. However, because the typical turn-on and turn-off procedures of a HP can cover more than one but not more than two sampling intervals, we consider power changes over two subsequent intervals computed as
\begin{equation}
	\Delta_2 p_k:=p_k-p_{k-2},\ k=2,\dots,N,
	\label{eq:delta2}
\end{equation}
and illustrated in the bottom plot in \fig\ref{fig:powerSteps}. The local extrema of the time series ${\Delta_2 p:=\{\Delta_2 p_k\}_{k=2}^N}$ are marked by red circles, and we summarize their indices by ${\set{I}_{\ts{extr}}\subset\{2,\dots,N\}}$. If the HP is the dominant load, these extrema can indicate changes of the operational state of the HP. The $\Delta_2 p$ data is expected to exhibit three clusters: a cluster centered around zero comprises local extrema arising from small fluctuations of the aggregate load, and a cluster with positive and one with negative mean that correspond to changes in the load caused by switch-ons  and -offs of the HP, respectively. We apply the $k$-means algorithm to group the data into these clusters, and denote by ${\set{I}_{\ts{on}},\set{I}_{\ts{off}}\subset\set{I}_{\ts{extr}}}$ the set of indices of $\Delta_2 p_k$-values that belong to the switch-on and -off clusters, respectively. The thresholds 
\begin{align}
\begin{split}
	\Delta_{2,\ts{on}} &:= \ts{quantile}_{0.05}(\Delta_2 p_k,\ k\in\set{I}_{\ts{on}}),\\
	\Delta_{2,\ts{off}} &:= \ts{quantile}_{0.95}(\Delta_2 p_k,\ k\in\set{I}_{\ts{off}}).
\end{split}
\end{align}
are used to process the time series $\{p_k\}_{k=0}^N$ and identify a switch-off if ${\Delta_2p_k\leq\Delta_{2,\ts{off}}}$ and a switch-on if ${\Delta_2p_k\geq\Delta_{2,\ts{on}}}$. We keep track of the operational state of the HP to enforce a series of strictly alternating switch-ons and -offs. In addition, the minimum on- and off-duration of the HP are taken into account. The exact procedure is discussed in our previous work \cite{Mueller2017}. If a switching event is detected at time $k$, the index $k$ is stored in the set $\set{I}_{\ts{switch-on}}$ or $\set{I}_{\ts{switch-off}}$, respectively, and an estimate of the exact switching time $\hat{t}_{\ts{switch}}$ is computed as
\begin{equation}
	\hat{t}_{\ts{switch}} = t_s(k-1) + t_s\left(\frac{p_{k}-p_{k-1}}{p_{k}-p_{k-2}}\right)
	\label{eq:switchTime}
\end{equation}
and stored as an estimated switch-on or switch-off time $\hat{t}^i_{\ts{on}}$ or $\hat{t}^i_{\ts{off}}$, respectively. Finally, estimates of the on- and off-durations of every heating cycle $i$ are computed as
\begin{align}
\hat{d}^i_{\ts{on}} &= \hat{t}^{i+1}_{\ts{off}}-\hat{t}^{i}_{\ts{on}} & & \ts{and} &
\hat{d}^i_{\ts{off}} &= \hat{t}^{i}_{\ts{on}}-\hat{t}^{i}_{\ts{off}}.
\end{align}
The results of the HP state estimation are shown in \fig\ref{fig:powerSteps}, where the shaded gray areas denote time periods during which the HP is believed to be running. 

An estimate of the HP's rated power is computed as
\begin{align}
\begin{split}
	\hat{p}_r =&\,(\ts{median}(\Delta_2 p_k,\,k\in\set{I}_{\ts{switch-on}})\\
	&\ -\ts{median}(\Delta_2 p_k,\,k\in\set{I}_{\ts{switch-off}}))/2.
	\label{eq:estRatedPower}
\end{split}
\end{align}

\subsection{Identification of the loss rate}
\label{ss:identLossRate}
\begin{figure}[tb]
\setlength{\abovecaptionskip}{-5pt}
\centering
\psfrag{xl}[Bc][Bc][1.3]{Outdoor air temperature $\Tout$ $\degc$}
\psfrag{yl1}[Bc][Bc][1.3]{$\bar{r}_l(\Tout)$ h$^{-1}$}
\psfrag{yl2}[Bc][Bc][1.3]{$\bar{r}_c(\Tout)$ h$^{-1}$}
\psfrag{yl3}[Bc][Bc][1.3]{$d_c(\Tout)$}
\resizebox{0.5\textwidth}{!}{
\includegraphics[]{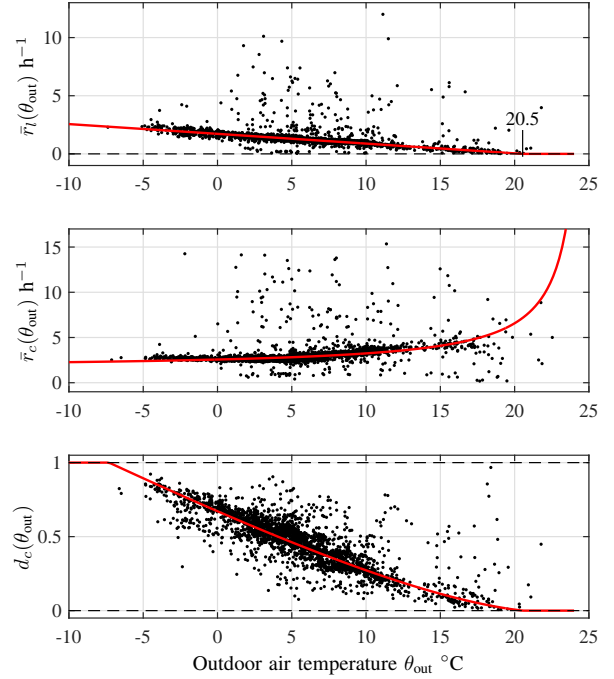}
}
\caption{Normalized loss rate $\bar{r}_l$ (top), charge rate $\bar{r}_c$ (middle), and corresponding duty cycle $d_c$ (bottom) versus outdoor air temperature. The solid lines represent the robust least-squares fits according to \eqref{eq:lossRateFcn} and \eqref{eq:chargeRateFcn}, respectively.}
\label{fig:ratesFit}
\end{figure}

The loss rate $\bar{r}_l^i$ describes the rate at which the normalized energy level $x(t)$ decreases over time. Its inverse equals the time required for the building to cool down from its upper energy bound $e_{\max}$ to its lower bound $e_{\min}$. The loss rate depends on different factors. However, here we only consider its dependency on the outdoor air temperature $\Tout$ by means of the piece-wise affine relationship 
\begin{equation}
	\bar{r}_l(\Tout)=\max(0,\,a_l\Tout+b_l),
	\label{eq:lossRateFcn}
\end{equation}
with parameters ${a_l,b_l\in\R{}}$. The temperature $\theta\opt_{\ts{out}}$, for which ${\bar{r}_l(\theta\opt_{\ts{out}})=0}$, is referred to as the \emph{zero-loss} temperature and can serve as an estimate of the building's nominal indoor air temperature. 
The top plot in \fig\ref{fig:ratesFit} shows the normalized loss rate versus the mean outdoor air temperature for all off-periods detected between September 1 2016 and March 15 2017. The red line depicts the robust least-squares fit of \eqref{eq:lossRateFcn}, with parameters ${a_l=-0.084}$ ($\degc$h)$^{-1}$, ${b_l=1.722}$ h$^{-1}$, and a zero-loss temperature ${\theta\opt_{\ts{out}}=20.5\,\degc}$. 

\subsection{Identification of the charge rate}
\label{ss:identChargeRate}
The charge rate $\bar{r}_c^i$ describes the rate at which the HP could increase the energy level $x(t)$ if there was no energy loss to the outside. Its inverse equals the time required for the HP to heat up the building from its lower bound $e_{\min}$ to its upper bound $e_{\min}$ if there are no thermal losses. The charge rate depends predominantly on the HP's rated power $\pr$ and its COP $\eta$, which is influenced by the out- and indoor air temperatures, the heating circuit water temperature $\theta_{\ts{w}}$, and other factors, such as the run-time of the HP. The outflow temperature $\theta_{\ts{w}}$ is usually determined by the \emph{heat curve}. We assume the affine relationship ${\theta_{\ts{w}}(\theta_{\ts{out}})=a_c\theta_{\ts{out}}+b_c}$ and neglect any saturation effects. Thus, our COP model is
\begin{equation}
	\eta(\theta_{\ts{out}}) = \zeta_{HP}\left(\frac{\theta_{\ts{w}}(\theta_{\ts{out}})+273.15}{\theta_{\ts{w}}(\theta_{\ts{out}})-\theta_{\ts{out}}}\right),
	\label{eq:eta}
\end{equation}
with $\zeta_{HP}$ being the exergetic efficiency. The constant $273.15$ is required to convert values from $\degc$ to K. Definition \eqref{eq:chargeRate} together with \eqref{eq:eta} suggest that the relationship between $\bar{r}_c$ and $\Tout$ is of the form
\begin{equation}
	\bar{r}_c(\Tout)=c_c\left(\frac{a_c\bar{\Tout}+b_c+273.15}{(a_c-1)\Tout+b_c}\right),
	\label{eq:chargeRateFcn}
\end{equation}
with parameters ${a_c,\,b_c,\,c_c\in\R{}}$. The red line in the middle plot in \fig\ref{fig:ratesFit} depicts the robust nonlinear least-squares fit of \eqref{eq:chargeRateFcn} with parameter values ${a_c=-17.85}$, ${b_c=473.26\,\degc}$, and ${c_c=1.6262}$ h$^{-1}$.

\subsection{Identification of the duty cycle}
\label{ss:identDutyCycle}

The duty cycle of the HP during the heating cycle $i$ is
\begin{equation}
	d_c^i = d^i_{\ts{on}}/(d^i_{\ts{on}}+d^i_{\ts{off}}).
	\label{eq:defDC}
\end{equation}
According to \eqref{eq:defNormRates}, the right-hand side of \eqref{eq:defDC} equals $\bar{r}_l^i/\bar{r}_c^i$. Thus, the duty cycle is given by the ratio ${\bar{r}_l(\Tout)/\bar{r}_c(\Tout)}$ and its inherent limitation to the interval $[0,1]$:
\begin{equation}
	d_c(\Tout)=\min(1,\,\max(0,\,\bar{r}_l(\Tout)/\bar{r}_c(\Tout))).
	\label{eq:dcFcn}
\end{equation}
The red line in the bottom plot in \fig\ref{fig:ratesFit} depicts the duty cycle of the HP according to \eqref{eq:dcFcn}. It takes the value 0 for ${\Tout\geq 20.5\,\degc}$ and 1 for ${\Tout\leq -7.4\,\degc}$.

\subsection{Flexibility level}
\label{ss:flexLevel}

Under undisturbed operating conditions, the TC of the HP keeps the indoor air temperature $\Tin$ within a range $[\theta_{\ts{in,min}},\theta_{\ts{in,max}}]$ that is comfortable for the building inhabitants. The width of this range directly affects the potential of the system to provide DR services because it limits the maximum on- and off-durations of the HP. 
In most cases, however, the original temperature range is chosen to be rather conservative, and the inhabitants accept wider ranges if they are awarded for it. We introduce a \emph{flexibility level} as a simple means for building owners to express their willingness to accept wider admissible indoor temperature ranges. Here we use a numeric flexibility level ${f\geq 1}$ that defines the maximum off-duration for the HP as
\begin{equation}
	d_{\ts{off,max}}(\Tout) := f d_{\ts{off}}(\Tout) = f/\bar{r}_l(\Tout).
	\label{eq:maxOffTime}
\end{equation}
The AG in charge of controlling the HP must respect this limit. Because in our setup the AG does not know how long a HP has been off at the time it is throttled, the maximum allowed throttle duration $T_{\max}$ is given as 
\begin{equation}
	T_{\max}(\Tout)=(f-1)/\bar{r}_l(\Tout).
	\label{eq:maxThrottleTime}
\end{equation}
That is, the AG is not allowed to throttle the HP if ${f=1}$ and the original admissible temperature range is maintained. Values of ${f>1}$ allow the AG to throttle the HP and can lead to indoor air temperatures below $\theta_{\ts{in,min}}$. The concept of the flexibility factor can be easily generalized to define the maximum on-duration of a HP in those cases where the HP can be fully controlled (in contrast to our combination of a TC and a throttling mechanism). Moreover, categorical values such as \{low, medium, high\} can be offered to end-customers and then mapped to numeric values by the AG.

\section{Population Modeling}
\label{s:popModeling}

In this section, we characterize the aggregate DR behavior of a population of $H$ buildings indexed by ${h\in\set{H}:=\{1\dots,H\}}$ by means of their aggregate load reduction potential and the corresponding rebound.


\subsection{Expected load reduction}
\begin{figure}[bt]
\setlength{\abovecaptionskip}{-2pt}
\centering
\psfrag{xl}[Bc][Bc][1.3]{$\Tout$ $\degc$}
\psfrag{yl}[Bc][Bc][1.3]{$\hat{p}_{\ts{red}}\ind{\set{H}}(\Tout)$ kW}
\resizebox{0.5\textwidth}{!}{
\includegraphics{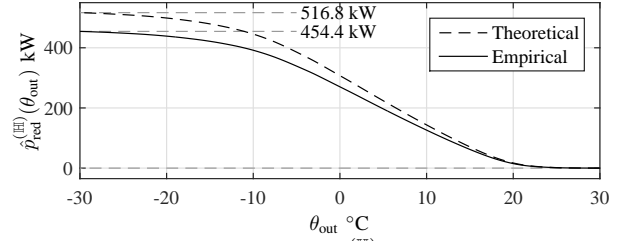}
}
\caption{Expected total load reduction $\hat{p}_{\ts{red}}\ind{\set{H}}(\Tout)$ that can be achieved by throttling all the 209 identified HPs simultaneously for different outdoor air temperatures $\Tout$.}
\label{fig:aggMaxLoadRed}
\end{figure}

Because SM data can be delayed by several hours, \cf\ref{s:expSetup}, the current state of a HP cannot be estimated by methods such as \cite{Vrettos2014b}. Instead, we consider the binary operational state of HP $h$ as a random variable whose probability of being ON equals its duty-cycle $d_c\ind{h}(\Tout)$. Thus, if a throttling command is sent to that HP, the expected load reduction is
\begin{equation}
	\hat{p}_{\ts{red}}\ind{h}(\Tout) = \hat{p}_r\ind{h}\hat{d}_c\ind{h}(\Tout).
\end{equation}
Experiments have shown that on average about 12\% of all HPs fail to respond to our throttling commands. Thus, we include the empirical success rate ${\sigma_{\ts{red}}=0.88}$ when computing the total expected load reduction as
\begin{equation}
	\hat{p}_{\ts{red}}\ind{\set{H}}(\Tout) = \sigma_{\ts{red}}\sum\limits_{h\in\set{H}}\hat{p}_{\ts{red}}\ind{h}(\Tout).
	\label{eq:maxAggLoadRed}
\end{equation}
\fig\ref{fig:aggMaxLoadRed} illustrates the theoretic and empirical total load reduction that can be achieved by our population of ${H=209}$ buildings for different outdoor air temperatures. A total load reduction of $454.4$ kW is reached at -30 $\degc$.

\begin{figure}[tb]
\setlength{\abovecaptionskip}{-5pt}
\setlength{\belowcaptionskip}{-15pt}
\centering
\psfrag{xl}[Bc][Bc][1.3]{$\Tout$ $\degc$}
\psfrag{yl}[Bc][Bc][1.3]{T h}
\psfrag{zl}[Bc][Bc][1.3]{$\hat{p}_{\ts{red}}\ind{\set{H}}(\Tout,T)$ kW}
\resizebox{0.5\textwidth}{!}{
\includegraphics{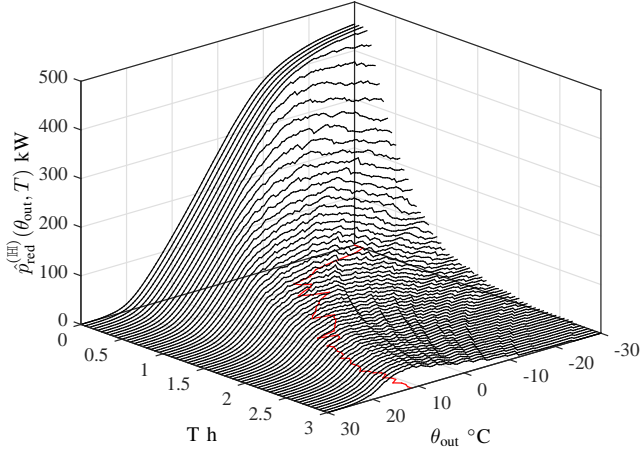}
}
\caption{Maximum load reduction $\hat{p}_{\ts{red}}\ind{\set{H}}(\Tout,T)$ expected from our population of 209 HPs for different outdoor air temperatures $\Tout$ and throttling durations $T$ with flexibility factors ${f\ind{h}=4}$, ${h\in\set{H}}$. The red line depicts the outdoor temperatures at which the largest load reduction is expected for a given throttling duration.}
\label{fig:expResponse}
\end{figure}

Let ${\set{I}_T(\Tout)\subseteq\set{H}}$ denote the set of HPs that can be throttled for a duration $T$ or longer for a given temperature $\Tout$, \ie,
\begin{equation}
	\set{I}_T(\Tout):=\{h\in\set{H}:\,T_{\max}\ind{h}(\Tout)\geq T\}.
	\label{eq:throttleSet}
\end{equation}
We characterize the aggregate load reduction potential of population $\set{H}$ by computing the maximum expected aggregate load reduction \eqref{eq:maxAggLoadRed} for different outdoor air temperatures $\Tout$ and throttling durations $T$ as 
\begin{equation}
	\hat{p}_{\ts{red}}\ind{\set{H}}(\Tout,T) = \sigma_{\ts{red}}\sum\limits_{h\in\set{I}_T(\Tout)}\hat{p}_{\ts{red}}\ind{h}(\Tout).
	\label{eq:aggLoadRed}
\end{equation}
The values of \eqref{eq:aggLoadRed} obtained for our population are shown in \fig\ref{fig:expResponse} for flexibility factors ${f\ind{h}=4}$, ${h\in\set{H}}$. The achievable load reduction decreases for longer throttling times because the number of HPs that can be throttled for the full duration $T$ declines. The gradual decline visible in \fig\ref{fig:expRebound} is a result of the heterogeneity of our population of systems with regard to the maximum throttle times \eqref{eq:maxThrottleTime}. With decreasing $\Tout$, the load reduction potential grows because the probability of a HP being ON increases according to its duty-cycle, \cf\fig\ref{fig:ratesFit}. The load reduction curves for a constant throttling duration do not increase monotonically because the set of HPs admissible for throttling, $\set{I}_T(\Tout)$, shrinks as $\Tout$ decreases. The red line depicts the outdoor temperatures at which the largest load reduction is expected for a given throttling duration.

\subsection{Expected rebound}
\begin{figure}[tb]
\setlength{\abovecaptionskip}{-5pt}
\setlength{\belowcaptionskip}{-15pt}
\centering
\psfrag{xl}[Bc][Bc][1.3]{$\Tout$ $\degc$}
\psfrag{yl}[Bc][Bc][1.3]{T h}
\psfrag{zl}[Bc][Bc][1.3]{$\hat{p}_{\ts{reb}}\ind{\set{H}}(\Tout,T)$ kW}
\resizebox{0.5\textwidth}{!}{
\includegraphics{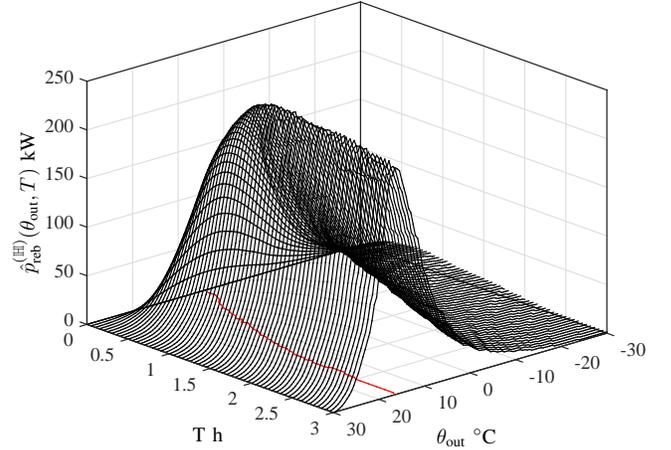}
}
\caption{Rebound power $\hat{p}_{\ts{reb}}\ind{\set{H}}(\Tout,T)$ expected from our population of 209 HPs for different outdoor air temperatures $\Tout$ and throttling durations $T$ with flexibility factors ${f\ind{h}=4}$, ${h\in\set{H}}$. The red line depicts the outdoor temperatures at which the largest rebound power is expected for a given throttling duration. Peak rebound values lie in the range of 230--247 kW.}
\label{fig:expRebound}
\end{figure}

After releasing the throttling signal of a HP at the end time $t_{\ts{end}}$ of a throttling period of duration $T$, the TC turns the HP on if the energy level of the building has fallen below its lower bound, \ie, ${x(t_{\ts{end}})\leq 0}$. The probability of this happening is
\begin{align}
	P(x(t_{\ts{end}})\leq 0) &= P(x(t_{\ts{end}}-T)\leq T\bar{r}_l\ind{h}(\Tout))\\
	&= T\bar{r}_l\ind{h}(\Tout),
	\label{eq:probRebound}
\end{align}
given the linear state dynamics \eqref{eq:socDyn}. This means that at time $t_{\ts{end}}$ the HP turns on with probability $T\bar{r}_l\ind{h}(\Tout)$ and consumes $\hat{p}_r\ind{h}$ units of power. If the HP had not been throttled, its expected average power consumption would be $d_c\ind{h}(\Tout)\hat{p}_r\ind{h}$. We define the expected rebound power $\hat{p}_{\ts{reb}}\ind{h}$ as the HP's excess consumption following a throttling period, \ie,
\begin{equation}
	\hat{p}_{\ts{reb}}\ind{h}(\Tout):=T\bar{r}_l\ind{h}(\Tout)(1-d_c\ind{h}(\Tout))\hat{p}_r\ind{h}.
	\label{eq:expRebPower}
\end{equation}
Similar to \eqref{eq:aggLoadRed}, we take an empirical success rate $\sigma_{\ts{reb}}$ into account when computing the aggregate rebound power expected from the entire population $\set{H}$ as
\begin{equation}
	\hat{p}_{\ts{reb}}\ind{\set{H}}(\Tout,T) = \sigma_{\ts{reb}}\sum\limits_{h\in\set{I}_T(\Tout)}\hat{p}_{\ts{reb}}\ind{h}(\Tout)
	\label{eq:aggRebound}
\end{equation}
which is shown in \fig\ref{fig:expRebound} for flexibility factors ${f\ind{h}=4}$, ${h\in\set{H}}$. For short throttling durations, the rebound grows with $T$ because longer throttling durations increase the probability of a rebound according to \eqref{eq:probRebound}. However, as $T$ grows further, fewer HPs can be throttled, \cf\eqref{eq:throttleSet}, and the rebound decreases again. For low $\Tout$, the rebound approaches zero because the HPs are ON most of the time, \ie, their duty-cycles approach 1. Thus, the rebound diminishes in accordance with \eqref{eq:expRebPower}. For high $\Tout$, in contrast, the expected rebound is small because of the diminishing probability \eqref{eq:probRebound} that the system reaches its lower state limit and triggers a rebound. Peak rebound values lie in the range of 230--247 kW for ${T>1}$ h and $\Tout$ between 11--16 $\degc$, see red line in \fig\ref{fig:expRebound}.

\section{Experimental Results}
\label{s:results}



\subsection{Load reduction}
We considered a population of 322 buildings with HP installations and conducted more than 70 DR experiments during which the HPs of at most 209 buildings were throttled for different durations and outdoor temperatures. In each experiment, we denote by $\set{I}_{\ts{throttle}}$ the set of participating systems, and refer to the rest as the reference group, $\set{I}_{\ts{ref}}$.

\begin{figure}[tb]
\setlength{\abovecaptionskip}{-5pt}
\centering
\psfrag{xl2}[Bc][Bc][1.3]{Time}
\psfrag{yl1}[Bc][Bc][1.3]{Power kW}
\psfrag{yl2}[Bc][Bc][1.3]{Power deviation kW}
\psfrag{red}[Bc][Bc][1.3]{Load reduction}
\psfrag{reb}[Bl][Bl][1.3]{Rebound}
\resizebox{0.5\textwidth}{!}{
\includegraphics{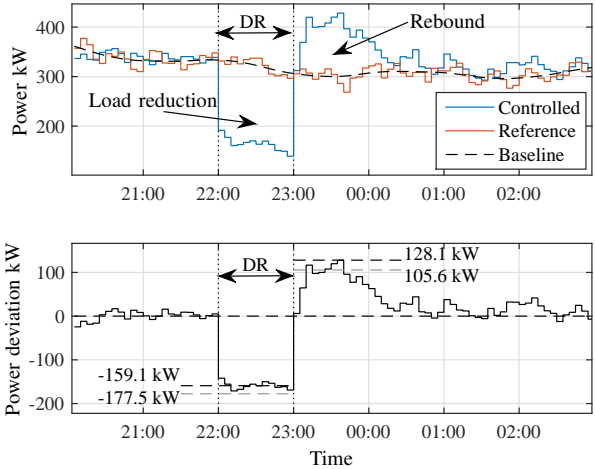}
}
\caption{Results from a demand response experiment involving 141 buildings. Top: Aggregate power consumption of the controlled buildings (blue), the scaled consumption of the reference group (red), and the fitted baseline (dashed black). A load reduction service is delivered between 22.00 h and 23.00 h followed by a rebound. Bottom: The load reduction and rebound are defined as the deviation of the power consumption of the controlled group from the baseline.}
\label{fig:drTest1}
\end{figure}

\fig\ref{fig:drTest1} shows a typical load reduction experiment: At 22.00 h, a throttling signal was sent to 141 out of 322 HPs and released one hour later. The remaining systems were used as the reference group. The expected aggregate load reduction ${\hat{p}_{\ts{red}}\ind{agg}(\hat{\theta}_{\ts{out}},T)=177.5}$ kW and the expected peak rebound power ${\hat{p}_{\ts{reb}}\ind{agg}(\hat{\theta}_{\ts{out}},T)=105.6}$ kW were computed in advance via \eqref{eq:aggLoadRed} and \eqref{eq:aggRebound}, respectively, with ${T=1}$ h and forecast ${\hat{\theta}_{\ts{out}}=0.1\,\degc}$. The solid blue line in the top plot depicts the aggregate consumption of the controlled buildings measured by their SMs on the timescale of 5 min. To make this consumption comparable to that of the reference group, the reference consumption was scaled so as to optimally match (in the sense of least squares) that of the controlled group during the 8 h preceding the throttling period. The scaled reference consumption is shown as a solid red line. A smoothing spline (dashed black) was then fitted to the scaled reference consumption data. The spline serves as the baseline relative to which the DR of the controlled group is measured, see bottom plot in \fig\ref{fig:drTest1}. The data show that the majority of the HPs turns off within the first 5 min of the throttling period. An average load reduction of 159.1 kW was achieved, which, compared with the predicted 177.5 kW, amounts to a load reduction prediction error of 11.6\%. After the throttling signal is released, most HPs turn back on to restore their nominal energy levels. This synchronization leads to a rebound \eqref{eq:expRebPower} with a peak at 128.1 kW. Compared with the predicted 105.6 kW, the rebound power prediction error amounts to 17.5\%.

\subsection{Rebound damping}
\begin{figure}[tb]
\setlength{\abovecaptionskip}{-5pt}
\centering
\psfrag{xl2}[Bc][Bc][1.3]{Time}
\psfrag{yl1}[Bc][Bc][1.3]{Power deviation kW}
\psfrag{yl2}[Bc][Bc][1.3]{Power deviation kW}
\resizebox{0.5\textwidth}{!}{
\includegraphics{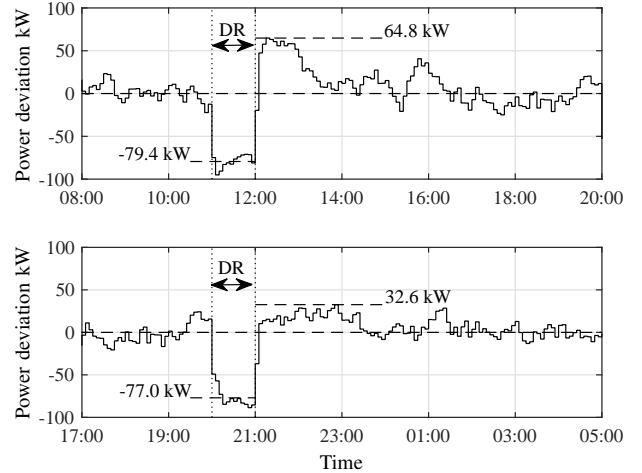}
}
\caption{Results from two 1-h load reduction experiments involving the same group of 74 buildings. Releasing the throttling signal simultaneously results in a pronounced rebound (top). The rebound damping strategy \eqref{eq:relTime} can be used to significantly reduce the rebound by spreading out the throttling release times of individual HPs (bottom).}
\label{fig:reboundDamping}
\end{figure}

Our results suggest that load reductions of thermostatically controlled loads are typically followed by a rebound period, during which the aggregate load is higher than usual as devices are trying to restore nominal conditions as quickly as possible, see top plots in Figures \ref{fig:drTest1} and \ref{fig:reboundDamping}. Different strategies have been proposed \cite{Motegi2007,Wrinch2012} to avoid the undesirable features of rebounds, such as sharp load ramps and high demand peaks. We introduce a rebound damping strategy similar to the \emph{sequential equipment recovery} described in \cite{Motegi2007}, in which the throttling signals of individual HPs are released sequentially rather than simultaneously to avoid synchronized switch-ons of HPs. Consider a group of HPs ${\set{I}_T\subseteq\set{H}}$ that have been throttled during the time interval $[t_{\ts{start}},t_{\ts{end}}]$. Instead of releasing the throttling signals for all HPs simultaneously at time $t_{\ts{end}}$, the throttling signal applied to device ${h\in\set{I}_T}$ is released at time
\begin{equation}
	t_{\ts{rel}}\ind{h} = \min(t_{\ts{end}}+\Delta T,\,t_{\ts{start}}+T_{\max}\ind{h}(\Tout)),
	\label{eq:relTime}
\end{equation}
where the parameter ${\Delta T\geq 0}$ bounds the time interval over which the individual release times can be dispersed. Choosing $\set{I}_T$ according to \eqref{eq:throttleSet} guarantees that ${t_{\ts{rel}}\ind{h}\geq t_{\ts{end}}=\Delta T}$.
\fig\ref{fig:reboundDamping} compares two load reduction experiments involving the same group of 74 buildings and HPs. In the experiment shown in the top plot, the throttling signals are released simultaneously at 12.00 h, \ie, ${\Delta T=0}$, which results in a sharp load ramp and a peak rebound of 64.8 kW. In the second experiment, shown in the bottom plot, ${\Delta T = 45}$ min was used to spread the individual release times \eqref{eq:relTime}. The rebound is reduced to values below 32.6 kW, which amounts to a peak rebound damping of 50\%. Our rebound damping strategy benefits from the heterogeneity among the systems: a wide range of $T_{\max}\ind{h}(\Tout)$ values in \eqref{eq:relTime} yields release times that are well spread out and that result in significant rebound damping. This highlights the importance of identifying the systems individually and taking into account their diversity. 
%

\subsection{Impact and accuracy of load reduction}
\begin{figure}[!b]
\setlength{\abovecaptionskip}{-2pt}
\centering
\psfrag{xl}[Bc][Bc][1.3]{$\Tout$ $\degc$}
\psfrag{yl}[Bc][Bc][1.3]{$p_{\ts{red}}\ind{\set{H}}$ \%}
\resizebox{0.5\textwidth}{!}{
\includegraphics{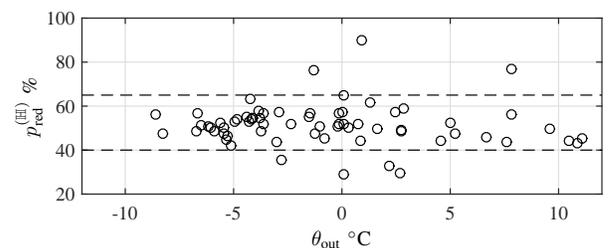}
}
\caption{Aggregate load reduction values $p_{\ts{red}}\ind{\set{H}}$ in percent of the aggregate load achieved in 67 experiments for different outdoor air temperatures $\Tout$. 57 experiments resulted in load reduction values in the range of 40--65\% of the aggregate load.}
\label{fig:loadRedPerc}
\end{figure}

Different groups of buildings have been used to achieve load reduction values in the range of 35--288 kW, which correspond to 29--90\% of the aggregate load, see \fig\ref{fig:loadRedPerc}. 57 of the 67 experiments resulted in an aggregate load reduction value in the range of 40--65\% over a broad air temperature range. Four experiments yielded load reduction values lower below 40\%, and three experiments resulted in exceptionally high values (76--90\%). The throttling periods of these experiments coincided with periods of high photovoltaic generation, which covered a significant share of the buildings' aggregate load. Thus, the HPs accounted for most of the remaining load. 

The expected load reduction and rebound values computed via \eqref{eq:aggLoadRed} and \eqref{eq:aggRebound} were compared with the actual values, and the prediction accuracy was assessed by means of the absolute percentage error (APE), whose statistics are summarized in \tab\ref{t:errorStats}. The average load reduction can be predicted accurately with a median error of 6.7\%. Predicting the peak rebound power is challenging because it heavily depends on the degree of synchronization among the HPs when resuming operation.

The results show that a population of heterogeneous residential heating systems can provide significant amounts of precisely predictable load reduction services to the SO.

\begin{table}[tb]
\renewcommand{\arraystretch}{1.3}
\caption{Load reduction and rebound prediction error statistics} 
\label{t:errorStats} 
\centering
\begin{tabular}{rcc} \hline
   &  Avg. load reduction &  Peak rebound\\ \hline
$\ts{quantile}_{0.25}$(APE) & 4.3\% & 12.5\%\\
median(APE)        & 6.7\%  & 22.9\%\\
$\ts{quantile}_{0.75}$(APE) & 12.3\% & 40.5\%\\
\hline
\end{tabular} 
\end{table}

%
\IEEEpeerreviewmaketitle

\section{Conclusion \& Future work}
\label{s:conclusion}
This work presented the results from a large-scale demonstration of a demand response scheme involving a population of over 300 residential buildings with heat pump installations. It was shown how the energetic behavior and flexibility of individual systems can be identified autonomously based only on readily available measurement data, and how the aggregate demand response potential of the entire population of buildings can be quantified. Experimental results illustrated the effectiveness of the approach: the load reductions can be predicted precisely and amount to 40--65\% of the aggregate load, and the associated rebound can be damped efficiently.

Future work will investigate how to increase the flexibility factor of individual systems with the aim of maximizing the flexibility available without compromising user comfort. Further, the quantification of the rebound in terms of peak power, energy content, and duration will be studied in more detail. Finally, the reliability of the communication and control setup will be improved.


%


\section*{Acknowledgment}
The authors gratefully acknowledge the fruitful collaboration with their colleagues in the \textsc{Ecogrid2.0} project.


\ifCLASSOPTIONcaptionsoff
  \newpage
\fi

\bibliographystyle{IEEEtran}
\bibliography{../../../Papers/BibTeX/library}
\enlargethispage{-5in}

\end{document}